\documentclass[sn-mathphys,Numbered]{sn-jnl}

\usepackage{graphicx}%
\usepackage{multirow}%
\usepackage{amsmath,amssymb,amsfonts}%
\usepackage{amsthm}%
\usepackage{mathrsfs}%
\usepackage[title]{appendix}%
\usepackage{xcolor}%
\usepackage{textcomp}%
\usepackage{manyfoot}%

\graphicspath{{./pics/}}

\newcommand{\ket}[1]{|#1\rangle}
\newcommand{\bra}[1]{\langle #1|}
\usepackage{placeins}
\usepackage{physics}

\usepackage{tikz}
\usetikzlibrary{quantikz2}

\usepackage{graphicx}
\usepackage{subcaption}
\usepackage[figurename=FIG., justification=RaggedRight]{caption}
\usepackage{listofitems} 
\usetikzlibrary{arrows.meta} 
\usepackage[outline]{contour} 

\tikzset{>=latex} 
\usepackage{xcolor}
\colorlet{myblue}{blue!80!black}
\colorlet{myorange}{orange!100!black}
\colorlet{mydarkblue}{blue!40!black}
\tikzstyle{node}=[thick,circle,draw=myblue,minimum size=22,inner sep=0.5,outer sep=0.6]
\tikzstyle{node in}=[node,orange!20!black,draw=myorange!30!black,fill=myorange!50]
\tikzstyle{node hidden}=[node,blue!20!black,draw=myblue!30!black,fill=myblue!40]
\tikzstyle{node out}=[node,orange!20!black,draw=myorange!30!black,fill=myorange!50]
\tikzstyle{connect}=[thick,mydarkblue] 
\tikzstyle{connect arrow}=[-{Latex[length=4,width=3.5]},thick,mydarkblue,shorten <=0.5,shorten >=1]
\tikzset{ 
  node 1/.style={node in},
  node 2/.style={node hidden},
  node 3/.style={node out},
}
\def\nstyle{int(\lay<\Nnodlen?min(2,\lay):3)} 

\begin{document}

\title[Article Title]{Echo-evolution data generation for quantum error mitigation via neural networks}


\author[1]{\fnm{Danila} \sur{Babukhin}}\email{dv.babukhin@gmail.com}

\affil[1]{\orgname{Dukhov Research Institute of Automatics (VNIIA)}, \orgaddress{\city{Moscow}, \postcode{127055}, \country{Russia}}}


\abstract{
Neural networks provide a prospective tool for error mitigation in quantum simulation of physical systems. 
However, we need both noisy and noise-free data to train neural networks to mitigate errors in quantum computing results.
Here, we propose a physics-motivated method to generate training data for quantum error mitigation via neural networks, which does not require classical simulation and target circuit simplification.
In particular, we propose to use the echo evolution of a quantum system to collect noisy and noise-free data for training a neural network.
Under this method, the initial state evolves forward and backward in time, returning to the initial state at the end of evolution. When run on the noisy quantum processor, the resulting state will be influenced by with quantum noise accumulated during evolution. Having a vector of observable values of the initial (noise-free) state and the resulting (noisy) state allows us to compose training data for a neural network. We demonstrate that a feed-forward fully connected neural network trained on echo-evolution-generated data can correct results of forward-in-time evolution. 
Our findings can enhance the application of neural networks to error mitigation in quantum computing.}




\maketitle

\clearpage
\section{Introduction}
\label{Intro}

Development of quantum computers in recent years led to demonstrating genuinely nontrivial results \cite{Arute2019,Kim2023} on devices with up to hundreds of qubits and non-ideal components \cite{Preskill2018}. 
Despite this, fault-tolerant quantum computing with error-correcting codes is still beyond nowadays technological level. 
Until fault-tolerant quantum computing unfolds, there is an alternative solution - to support NISQ devices with quantum error mitigation (QEM)\cite{Cai2023}. Quantum error mitigation is an approach to reduce the effect of quantum noise in the results of quantum computing via additional data sampling and post-processing.
Up to date, several examples of QEM have been demonstrated  \cite{Temme2017,Cai2021,Huggins2021} and theoretical understanding of QEM has advanced \cite{Cai2023practical} to make QEM a convenient tool to support further development of quantum computing.

A data-driven approach to quantum error mitigation \cite{Czarnik2021,czarnik2022improving,Lowe2021,Zhukov2022,Lee2023} is a branch of QEM, which is a promising enhancement to the existing QEM toolbox. 
The data-driven approach to QEM includes gathering noisy and noisy-free data and fitting (training) an ansatz function to approximate a mapping between noisy and noisy-free data. The trained ansatz function is then used to post-process data from a noisy quantum device. To date, there are several works related to data-driven QEM.
In \cite{Czarnik2021}, the idea to use Clifford gates to generate classical data to provide ideal observables of a target circuit, is proposed. 
The data is then used to train a linear ansatz function to error mitigation. 
This method was demonstrated to be used with other QEM techniques to provide good error mitigation performance \cite{Lowe2021} and provides high flexibility in terms of combining with other error mitigation methods \cite{Bultrini2023_2}.
Another group of works demonstrates the use of non-linear ansatz functions - neural networks - in quantum error mitigation. There was demonstrated the capability to correct dynamics of many-body observables, corrupted with quantum gate errors \cite{Zhukov2022}, and the capability to mitigate measurement errors \cite{Lee2023}. 

Although the best machine learning approach to each case of QEM is unknown, a definite advantage of machine learning is a possibility to move computational overhead from quantum computing runtime to a training phase of a machine learning algorithm \cite{liao2023machinelearningpracticalquantum}. During quantum computing runtime, we have to run a device several times to construct a noiseless value estimator via non-machine-learning quantum error mitigation method. In contrast, we can apply a machine learning estimator directly to a single noisy value of a target quantum observable, thus running a quantum device only once. But in that case, we need to train a machine learning algorithm quantum error mitigation, and this is where additional computational overhead occurs. Even if machine learning does not eliminate computational overhead, moving it from quantum computing runtime allows speeding up useful computation of a NISQ device. As a result, machine learning QEM may promote application of NISQ devices to real-world problems.

Training neural networks requires gathering data set, which consist of corresponding noisy and noise-free observable values. For a problem of quantum dynamics simulation, our final target is a quantum state of the system at model time $t$. To have noise-free observable data on such a state requires us to simulate the dynamics classically - a generally infeasible task for systems in the regime of quantum advantage ($>$ 50 qubits). Thus, there is a need for a workaround method to generate noise-free data for quantum error mitigation.
A possible solution to this problem was proposed in \cite{Zhukov2022}. There, lower-depth versions of quantum circuits are used to generate data output of target (deep-depth) circuits. As the circuit in \cite{Zhukov2022} is a Trotter decomposition of the evolution operator, the depth of the circuit is controlled by a number of Trotter steps. 
Although being a working solution, this approach bounds the opportunity to simulate quantum dynamics for long times as a lower Trotter approximation leads to lower quantum simulation accuracy. 

In this paper, we propose a physics-motivated method to generate data for quantum error mitigation via neural networks, which does not require classical simulation and target circuit simplification. 
In particular, we demonstrate that the echo evolution - an evolution of a quantum system forward and backward in time - allows the production of noisy and noise-free data required to train a neural network. 
We show that a neural network trained on such data can mitigate the effect of quantum noise on the results of forward-in-time evolution, which is the actual target of quantum simulation. 
To illustrate the proposed method, we simulate dynamics of a 2D spin system under the transverse-field Ising Hamiltonian. We show that, for realistic gate noise, the neural network, trained on echo-evolution-generated data, mitigates effect of gate noise in forward-in-time dynamics results.

This paper is organized as follows. 
We provide background on the transverse-field Ising model in Sec.\ref{subsection2A}. 
We formulate a method of data generation via echo-evolution in Sec.\ref{EchoMethod}. 
We provide results on QEM via a feed-forward neural network, trained on data from echo evolution of a 6-spin system, in Sec.\ref{Forward_data_result}. 
We provide a case-study analysis of the hidden layer width of our neural network in Sec.\ref{AnalysisWidth} and show that the quality of error mitigation saturates for a number of neurons in the hidden layer, approximately equal to the size of simulated system data vectors. 
We draw a conclusion in Sec.\ref{Conclusion}

\section{Dynamics of the transverse field Ising model}
\label{subsection2A}

One of the main problems, which has an advantage of using quantum computing, is the simulation of quantum system dynamics. As was initially proposed by Feynmann \cite{Feynmann1982} and Manin \cite{Manin1980}, quantum computing allows working with the whole basis of the Hilbert space of the quantum system, which is exponential in the number of particles in the system. Among others, simulating the dynamics of quantum spin systems is one of the most important parts of quantum computing many-body physics \cite{Smith2019}. The main model of spin many-body physics is the transverse-field Ising model \cite{mbeng2020}. This model allows investigating various many-body phenomena \cite{Smith2019, Mi2021,Chen2022}.
A hamiltonian of the transverse field Ising model is
\begin{equation}
    \label{Ising_hamiltonian}
    H = -h\sum_{i=1}^{N}\sigma^{X}_{i} - J\sum_{ij}^{N}\sigma^{Z}_{i}\sigma^{Z}_{j}
\end{equation}
where $h > 0$ is on-site energy of a single spin and $J > 0$ is interaction energy between coupled spins, and $\sigma^{Z}_{i}$, $\sigma^{X}_{i}$ are Pauli matrices with eigenvalues $\pm 1$. The quantum spin system undergoes evolution under this hamiltonian from an initial state $\ket{\psi(0)}$ at time $t=0$ to a final state $\ket{\psi(t)}$ at arbitrary time $t$, and the two states are connected as
\begin{equation}
    \ket{\psi(t)} = e^{-iHt}\ket{\psi(0)}
\end{equation}
where $e^{-iHt}$ is an evolution operator. The dynamics of this spin system is straightforward to run on a quantum computer: every single spin maps on a single qubit, and the evolution operator consists of single- and two-qubit gates available on the hardware. A convenient way of constructing the evolution operator from quantum gates is via Trotter decomposition. For a hamiltonian with two non-commuting parts $H_{A}$ and $H_{B}$ ($[H_{A}, H_{B}] \neq 0$) the Trotter decomposition for $N_{tr}$ steps is 
\begin{equation}
    \label{Trotter}
    e^{it(H_{A} + H_{B})} \approx (e^{i\frac{t}{N}H_{A}}e^{i\frac{t}{N}H_{B}})^{N_{tr}}
\end{equation}
with an error of the size $O(\frac{t^{2}}{N_{tr}})$. There are Trotter decompositions of the higher order \cite{Hatano2005}, which provide lower decomposition error and allow simulating dynamics for longer times.  

After evolving the initial state $\ket{\psi(0)}$ to the final state $\ket{\psi(t)}$, we usually measure a quantum observable $\mathcal{\hat{O}}$, which corresponds to a physical quantity in the modeled system. For the $N$-qubit system, the observable can be of the following form
\begin{equation}
    \label{observable_gen}
    \mathcal{\hat{O}} = \hat{O}_{1} \otimes \hat{O}_{2} \otimes \dots \otimes \hat{O}_{N}
\end{equation}
where $O_{j}$ is a single-qubit observable. In quantum computing, we usually have hardware-implemented access to measurements on a computational basis, which allows measuring operators of the form
\begin{equation}
    \mathcal{\hat{O}}_{i} = \ket{i_{1}}\bra{i_{1}} \otimes \ket{i_{2}}\bra{i_{2}} \otimes \dots \otimes \ket{i_{N}}\bra{i_{N}}
\end{equation}
for $i = 0, 1, ..., 2^{N}-1$ and binary representation $i = i_{1}i_{2}...i_{N}$. 
Using computational basis, we can measure qubit excitations via measurement
\begin{equation}
    \hat{n}^{j} = I \otimes ... \otimes \ket{1_{j}}\bra{1_{j}} \otimes ... \otimes I
\end{equation}
where we measure projection of the $j$-th qubit on $\ket{1}$ and sum over all other qubits outcomes. If we simulate a spin system using qubits, we can measure magnetization of individual spins via measuring qubit excitation $n^{j}$ and then calculating magnetization as $m^{j} = 2n^{j} - 1$. Finally, an average magnetization for a spin system is
\begin{equation}
    \label{average_m}
    M = \frac{1}{N}\sum_{j}m^{j}
\end{equation}


\section{Data generation via echo evolution}
\label{EchoMethod}

To use neural networks for quantum error mitigation, we need to collect data with noise-free observable values ($Y_{data}$). Generally, we want to use a quantum processor to run problems that are beyond the reach of classical computing. Thus, for noisy data from a NISQ device ($X_{data}$), we usually cannot simulate noise-free data ($Y_{data}$). 

There are possible solutions throughout the literature. For quantum measurement error mitigation, small depth circuits with single-qubit rotations allow constructing states with classically-predictable counts statistic \cite{Kim2020,Lee2023}. For the QAOA problem, using Clifford-gate circuits demonstrated successful error mitigation for ground state energy of a spin system \cite{Czarnik2021}. Finally, using data from forward-in-time dynamics with fewer deep circuits allowed mitigating errors in observable dynamics in deep circuits in a problem of quantum simulation of spin system dynamics \cite{Zhukov2022}. 

This work concentrates on the last mentioned problem - simulation of quantum dynamics. In \cite{Zhukov2022}, the data generation procedure produces training data using shallow-depth circuits of the target system evolution. As a quasi-ideal data, a decomposition of the evolution operator (\ref{Trotter}) with $N_{1}$ Trotter steps is used, with $N_{1}$ small enough to make the gate error effect negligible. Then, noisy data is generated by choosing a target number of Trotter steps $N_{2}$ and adding ``delay'' circuits to make quasi-ideal circuits of depth $N_{1}$ to make the noise effect high enough for the depth level $N_{2}$. This procedure generates a training data set of noisy observables $X_{data}$ and noise-free observables $Y_{data}$ to train a feed-forward neural network to mitigate the noise effect. 

The main drawback of the described method is a restriction on the quasi-ideal circuits depth. A limited number of Trotter steps is allowed to make the data produced with quasi-ideal circuits approximately noise-free. This limitation restricts the user to generating observable data only for small-time dynamics since the accuracy of simulated dynamics depends on the accuracy of Trotter decomposition (\ref{Trotter}): the fewer Trotter steps are used, the less accurate is the resulting dynamics.

\begin{figure}
    \centering
     \begin{subfigure}[b]{0.99\textwidth}
        \centering
        \includegraphics[width=0.8\linewidth] 
             {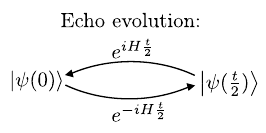}
        \caption{}
        \label{fig:fig1a}
     \end{subfigure}
     \hfill
     \begin{subfigure}[b]{0.99\textwidth}
        \centering
        \includegraphics[width=0.8\linewidth] 
             {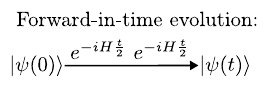}
         \caption{}
         \label{fig:fig1b}
     \end{subfigure}
     \caption{\textbf{The schematic description of the echo evolution and forward-in-time evolution.} \textbf{a} During echo evolution, a system evolves forward and backward during overall time $t$. In case of no noise during evolution, the system will return to the initial state. \textbf{b} During the forward-in-time evolution, the system evolves during overall time $t$ to some final state.}
    \label{fig:echo_forward_evolution}
\end{figure}

We may find a solution from physical insights of quantum systems evolution. In general, given an initial state $\ket{\psi_{init}}$, it is only possible to find the value of observable $O$ on the final state $\ket{\psi_{final}}$ (after evolution) via running the system evolution 
\begin{equation}
    \ket{\psi_{final}} = e^{-iHt}\ket{\psi_{init}}
\end{equation}
and then measuring an observable of interest
\begin{equation}
    \mathcal{O}(t) = \bra{\psi_{final}}\mathcal{\hat{O}}\ket{\psi_{final}}.
\end{equation}
However, there is a specific kind of evolution where knowing the initial state is enough to know the answer after evolution - the echo evolution (Fig.~\ref{fig:fig1a}). The echo evolution is the following:
\begin{equation}
    \ket{\psi_{final}} = e^{iH\frac{t}{2}}e^{-iH\frac{t}{2}}\ket{\psi_{init}} = \ket{\psi_{init}}
\end{equation}
The resulting observable value is then
\begin{equation}
    \label{Orelation}
    \mathcal{O}(t) = \bra{\psi_{final}}\mathcal{\hat{O}}\ket{\psi_{final}} = \bra{\psi_{init}}\mathcal{\hat{O}}\ket{\psi_{init}} = 
    \mathcal{O}(0).
\end{equation}
The relation (\ref{Orelation}) states that given an initial quantum state $\ket{\psi_{init}}$, the observable value after a unitary echo evolution is known beforehand. However, when we run the echo evolution on a noisy quantum device, the observable value after the evolution will be influenced by the overall quantum noise on a quantum device. In other words, we know a desirable outcome from a quantum device and the error-corrupted outcome from running the echo-evolution.


\begin{figure}[ht]
	\centering
	\includegraphics[width=0.9\linewidth] 
             {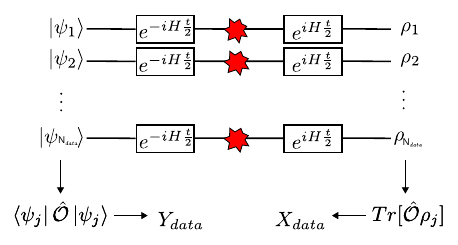}
	\caption{
\textbf{Data generation scheme.}
We prepare $N_{data}$ random initial states $\ket{\psi_{j}}, j = 1,\dots,N_{data}$. We measure the observable $\mathcal{\hat{O}}$ on each initial state to produce noise-free data. Then, every initial state is prepared again and is subject to echo-evolution under the transverse-field Ising hamiltonian (\ref{Ising_hamiltonian}). 
During the evolution, the system is subject to quantum noise, which we depict with a red star in the middle of each circuit line.
After evolution, the system ends in a state $\rho_{j}$, on which we measure the observable $\mathcal{\hat{O}}$ and obtain error-corrupted values.  
 }
	\label{fig:data_echo_evolution}
\end{figure}

The echo-type dynamics has many applications across physics. First examples are photon (or spin) echo phenomena \cite{allen1975optical}, where a sequence of $\frac{\pi}{2}$ and a $\pi$ pulses allows restoring coherence between atomic dipoles (or nuclear spins) and thus see an echo-radiation pulse as a result of restored collective evolution. These two phenomena were subjects of intense investigation in the second half of 20th century \cite{Hanh1950,Abella1966}. Next, the investigations on echo-dynamics in many-body systems lead to introduction of the Loschmidt echo \cite{Peres1984}. This is an evolution of a quantum system back and forth, with a perturbed hamiltonian, which allows investigating stability of system dynamics. It is initially introduced as a means to characterize emergence of chaotic behavior in quantum systems, and now this is an important theoretic tool in many fields of physics \cite{Wisniacki2012,Serbyn2017}. 
There are several examples of the application the echo evolution in quantum information processing. 
First, the Randomized Benchmarking protocol \cite{Emerson2005,Knill2008,Magesan2011} - a standard tool for evaluating the quality of quantum gates - is also based on a kind of echo evolution, when a sequence of random gates is applied to initial state and the ideal outcome (when no gate errors occur) is the same initial state. 
Next, back and forth evolution is a part of the recent algorithms for simulating quantum chemistry \cite{Barison2021,Bultrini2023}. One can use the uncomputation of a prepared and time-evolved step with a variational ansatz to obtain a compressed circuit representation of the ansatz time evolution and thus decrease the number of gates in the quantum chemical simulation, thus reducing noise influence on the calculation results. 

In this work, we use echo-evolution for the purpose of quantum error mitigation. The main goal of quantum error mitigation in the simulation of quantum dynamics is to mitigate noise influence on observables from forward-in-time dynamics (Fig.~\ref{fig:fig1b}). The data generation method provided here uses the same evolution operator of the quantum system, which produces the target (forward-in-time) evolution. If the depth of circuits, which generate echo and forward-in-time evolutions, is the same, then these evolutions are subject to the same level of noise. Thus, a neural network trained on an echo-evolution-generated data set can likely correct the effect of noise on the forward-in-time evolution outcome. As we demonstrate in the following section, this is indeed the case.

\section{Methods}
\label{AppendixA}

\subsection{Spin system evolution}
\label{AppendixA1}

To demonstrate the idea, we simulate the evolution of a 6-spin system under the transverse-field Ising hamiltonian (\ref{Ising_hamiltonian}). We choose parameters $h=1$ and $J = h/2$ and set the system a ladder topology (see Fig.~\ref{fig:qubit_layout}). We mapped every spin to a single qubit. We implemented the unitary evolution operator $e^{-iHt}$ (the operator in Fig.~\ref{fig:data_echo_evolution}) with noisy single- and two-qubit gates via Trotter decomposition (\ref{Trotter}) with $H_{A} = -h\sum_{i=1}^{N}\sigma_{i}^{X}$ and $H_{B} = -J\sum_{ij}^{N}\sigma_{i}^{Z}\sigma_{j}^{Z}$. 

For simulations, we chose a fixed number of Trotter steps such that the highest time point of the forward-in-time evolution is $T_{final} = \pi$. To prepare the training set via echo evolution, we use quantum circuits with the same number of Trotter steps to preserve the circuit depth. Unlike forward evolution, during echo evolution half of Trotter steps evolve the system backwards in time. Thus, during echo evolution the system evolves at most to the time $T_{echo} = \frac{T_{final}}{2} = \frac{\pi}{2}$ and then back to $t = 0$.

\begin{figure}[t!]
\begin{subfigure}[b]{0.99\textwidth}
        \centering
        \begin{tikzpicture}
        \node[scale=0.99] {
         \begin{quantikz}
             \lstick{$q_{0}$} & \gate{R_{X}} & \gate[2,style={fill=red!30}]{R_{ZZ}} & 
             \gate[3,style={fill=blue!30},label style={yshift=0.0cm}]{R_{ZZ}} & \qw & 
             \qw & \qw & 
             \qw & 
             \\
             \lstick{$q_{1}$} & \gate{R_{X}} & \qw & 
             \linethrough & \gate[3,style={fill=blue!30},label style={yshift=0.0cm}]{R_{ZZ}} & 
             \qw & \qw & 
             \qw & 
             \\
             \lstick{$q_{2}$} & \gate{R_{X}} & \gate[2,style={fill=red!30}]{R_{ZZ}} & 
             \qw & \linethrough & 
             \gate[3,style={fill=green!30},label style={yshift=0.0cm}]{R_{ZZ}} & \qw & 
             \qw & 
             \\
             \lstick{$q_{3}$} & \gate{R_{X}} & \qw & 
             \qw & \qw &  
             \linethrough & \gate[3,style={fill=green!30}]{R_{ZZ}} & 
             \qw & 
             \\
             \lstick{$q_{4}$} & \gate{R_{X}} & \gate[2,style={fill=red!30}]{R_{ZZ}} & 
             \qw & \qw & 
             \qw & \linethrough & 
             \qw & 
             \\
             \lstick{$q_{5}$} & \gate{R_{X}} & \qw & 
             \qw & \qw &
             \qw & \qw & 
             \qw & 
             \\
            \end{quantikz}
            };
        \end{tikzpicture}
        \caption{}
\end{subfigure}
\hfill
\begin{subfigure}[b]{0.99\textwidth}
        \centering
        \begin{tikzpicture}
        \node[scale=0.99] {
        \begin{quantikz}
            \lstick{$q_{i}$} & \gate[2]{R_{ZZ}} & \qw &
             \\
            \lstick{$q_{j}$} & & \qw &
        \end{quantikz}
        =
        \begin{quantikz}
            \lstick{$q_{i}$} & \ctrl{1} & \qw & \ctrl{1} & \qw
             \\
            \lstick{$q_{j}$} & \targ{} & \gate{R_{Z}(J_{ij}t)} & \targ{} & \qw
        \end{quantikz}
            };
        \end{tikzpicture}
        \caption{}
\end{subfigure}
	\caption{\textbf{(a):} A single Trotter step of evolution operator of the transverse field Ising hamiltonian. Gates are grouped to layers (colors) which can be simultaneously run on a quantum device (gates which do not have commonly used qubits). \textbf{(b)}: a decomposition of $R_{ZZ}$ operator to CNOT gates and a single-qubit rotation.}
	\label{fig:Trotter_step_decomposed}
\end{figure}

\subsection{Data generation}
\label{AppendixA2}

\begin{figure}
    \centering
     \begin{subfigure}[b]{0.35\textwidth}
        \centering
        \includegraphics[width=0.9\linewidth] 
             {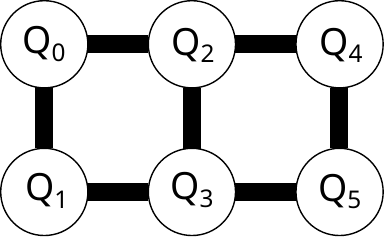}
        \vspace{1.25cm}
        \caption{Every single qubit corresponds to a single spin. Every two connected qubits can be subject to a two-qubit gate, which allows the implementation of interaction between two spins.}
        \label{fig:qubit_layout}
     \end{subfigure}
     \hfill
     \begin{subfigure}[b]{0.64\textwidth}
         \centering
             \begin{quantikz}
                 \lstick{$\ket{0_{0}}$} & \gate{U_{2}(\theta_{0}, \phi_{0})}
                 & \ctrl{1}      
                 & \qw & \ctrl{2} & \qw
                 & \qw & \qw & \\
                 \lstick{$\ket{0_{1}}$} & \gate{U_{2}(\theta_{1}, \phi_{1})}     
                 & \targ{}           
                 & \qw & \qw & \ctrl{2}
                 & \qw & \qw & \qw \\
                 \lstick{$\ket{0_{2}}$} & \gate{U_{2}(\theta_{2}, \phi_{2})}     
                 & \ctrl{1}          
                 & \qw & \targ{}
                 & \qw & \ctrl{2} & \qw & \qw  \\
                 \lstick{$\ket{0_{3}}$} & \gate{U_{2}(\theta_{3}, \phi_{3})}     
                 & \targ{}           
                 & \qw & \qw & \targ{}
                 & \qw & \ctrl{2} & \qw \\
                 \lstick{$\ket{0_{4}}$} & \gate{U_{2}(\theta_{4}, \phi_{4})}     
                 & \ctrl{1}          
                 & \qw & \qw
                 & \qw & \targ{} & \qw & \qw  \\
                 \lstick{$\ket{0_{5}}$} & \gate{U_{2}(\theta_{5}, \phi_{5})}     
                 & \targ{}            
                 & \qw & \qw 
                 & \qw & \qw  & \targ{} & \qw 
             \end{quantikz}
         \caption{First layer of single qubit rotation is parametrized with random angles $\theta_{j} = arccos(x)$, $x \in Uniform[-1,1]$ and $\phi_{j} \in Uniform[0,2\pi]$. CNOT gates are applied with probability $p = 0.2$ between every two connected qubits.}
         \label{fig:prepcirc}
     \end{subfigure}
     \caption{(a) A layout of a qubit system used in simulation in this work. (b) A quantum circuit implementing random initial states.}
    \label{fig:data_gen}
\end{figure}

To train a neural network, we gather data from random initial states for a 6-spin system. These states are generated using a circuit in Fig.~\ref{fig:prepcirc}.
Here, $\theta_{j} = arccos(x)$, $x \in Uniform[-1,1]$, $\phi_{j} \in Uniform[0,2\pi]$, and ``Random CNOTs'' corresponds to random CNOT gates, applied to every pair of layout-connected qubits (see Fig.~\ref{fig:qubit_layout}) with a probability $p = 0.2$. 
For every time point, the evolution operator consisted of $N_{tr}$ Trotter steps forward and $N_{tr}$ Trotter steps backward, with $N_{tr} = 10$, and the coupling map corresponding to qubit layout (\ref{fig:qubit_layout}). The overall gate count of $20$ Trotter steps includes 280 CNOT gates, 140 $R_{z}$ gates, and 120 $R_{x}$ gates.
We run simulation of echo evolution for $2400$ random initial states with $5$ time points $t \in [0, \frac{\pi}{8}, \frac{\pi}{4}, \frac{3\pi}{8}, \frac{\pi}{2}]$. Using several time points allows for the production of more data, as every time point gives one data vector. For $5$ time points, every random initial state produces $5$ data vectors, which differ by the mid-evolution state at time $t$. As a result, we obtain data from evolution with different times and amounts of entanglement generated by hamiltonian (\ref{Ising_hamiltonian}), and thus with noisy outcome data for these different entanglements. 
Data generation described above resulted in $12000$ pairs of noisy and noise-free observable vectors. 

We compose data vectors of single spin magnetizations and thus have data of the dimension of the number of spins $N$.
For each initial state, we produce a vector of single spin magnetizations from measurements done before and after echo-evolution.  We obtain pairs of noisy and noise-free magnetization vectors of the form
\begin{eqnarray}
    \label{magnetization_vect}
    \vec{m}_{noisy} = 
    \begin{pmatrix}
    \vspace{0.15cm}
    m^{0}_{noisy} \\
    \vspace{0.15cm}
    m^{1}_{noisy} \\
    \vspace{0.15cm}
    m^{2}_{noisy} \\
    \vspace{0.15cm}
    m^{3}_{noisy} \\
    \vspace{0.15cm}
    m^{4}_{noisy} \\
    \vspace{0.15cm}
    m^{5}_{noisy}     
    \end{pmatrix} 
    = 
    \begin{pmatrix}
    \vspace{0.15cm}
    2n^{0}_{noisy} - 1 \\
    \vspace{0.15cm}
    2n^{1}_{noisy} - 1 \\
    \vspace{0.15cm}
    2n^{2}_{noisy} - 1 \\
    \vspace{0.15cm}
    2n^{3}_{noisy} - 1 \\
    \vspace{0.15cm}
    2n^{4}_{noisy} - 1 \\
    \vspace{0.15cm}
    2n^{5}_{noisy} - 1     
    \end{pmatrix},
    \hspace{0.1cm}
    \vec{m}_{ideal} = 
    \begin{pmatrix}
    \vspace{0.15cm}
    m^{0}_{ideal} \\
    \vspace{0.15cm}
    m^{1}_{ideal} \\
    \vspace{0.15cm}
    m^{2}_{ideal} \\
    \vspace{0.15cm}
    m^{3}_{ideal} \\
    \vspace{0.15cm}
    m^{4}_{ideal} \\
    \vspace{0.15cm}
    m^{5}_{ideal}     
    \end{pmatrix} 
    = 
    \begin{pmatrix}
    \vspace{0.15cm}
    2n^{0}_{ideal} - 1 \\
    \vspace{0.15cm}
    2n^{1}_{ideal} - 1 \\
    \vspace{0.15cm}
    2n^{2}_{ideal} - 1 \\
    \vspace{0.15cm}
    2n^{3}_{ideal} - 1 \\
    \vspace{0.15cm}
    2n^{4}_{ideal} - 1 \\
    \vspace{0.15cm}
    2n^{5}_{ideal} - 1     
    \end{pmatrix}.
\end{eqnarray}
Here, $n_{ideal}^{i} (n_{noisy}^{i})$ is an excitation number of $i$-th qubit, measured before (after) the echo-evolution.
Correspondingly, $m_{ideal}^{i} (m_{noisy}^{i})$ is a magnetization of the $i$-th spin, computed from excitation number $n_{ideal}^{i} (n_{noisy}^{i})$. 

To test the performance of a trained neural network, we generate 100 random states and subject them to forward-in-time evolution from $t = 0$ to $t = \pi$. During the evolution, the system is subject to the same noise structure and level and with $N_{tr} = 20$ Trotter steps to have the same circuit depth (see information about the total number of gates above). 

\subsection{Error model}
\label{AppendixA3}
In this work, we focused on the mitigation of gate errors. In the current state of quantum computing hardware, gate errors represent one of the main sources of errors \cite{Kim2023} in large circuits, and most quantum error mitigation methods are first tested against this type of noise \cite{Cai2023}. 
As a quantum noise model, we used single- and two-qubit depolarizing errors
\begin{eqnarray}
    \label{noisemodel}
    \Phi^{depol}_{1q}(\rho) = (1 - q_{1})\rho + \frac{q_{1}}{2}I, \\
    \Phi^{depol}_{2q}(\rho) = (1 - q_{2})\rho + \frac{q_{2}}{4}I \otimes I
\end{eqnarray}
with noise intensities $q_{1} = 10^{-4}$ and $q_{2} = 0.01$ which correspond to state-of-the-art quantum computing capabilities \cite{Kim2023}. We do not consider SPAM errors and decoherence in this work to provide a proof-of-principle illustration of the proposed idea.

\subsection{Neural network structure, training and use}
\label{AppendixA4}
A neural network we use here is a multi-layer perceptron (see Fig.~\ref{fig:neural_network}). This neural network has $N_{input} = 6$ neurons in the input layer, a single hidden layer with $N_{hidden} = 200$ neurons, and the output layer with $N_{output} = 6$ neurons. The activation function of the hidden layer (denoted $\sigma_{1}$ in Section \ref{supervisedlearning}) is ReLU
\begin{equation}
    \sigma_{1}(x) = max(x, 0)
\end{equation}
and the activation function of the output layer (denoted $\sigma_{2}$ in Section \ref{supervisedlearning}) is Tanh
\begin{equation}
    \sigma_{2}(x) = \frac{e^{x} - e^{-x}}{e^{x} + e^{-x}}
\end{equation}
We use a mean square error (MSE) as a loss function
\begin{eqnarray}
    \label{MSE}
    R(\hat{W}, \hat{b}) = \frac{1}{N_{data}} \sum_{i=1}^{N_{data}}(f(\vec{x}_{i}, \hat{W}, \hat{b}) - \vec{y}_{i})^{2}
\end{eqnarray}
and minimize it using an Adam optimizer \cite{kingma2017adam} with parameters $lr = 3 * 10^{-4}$, $\beta_{1} = 0.9, \beta_{2} = 0.999$. We used a batch size of $80$ data vectors. We divide $12000$ data vectors, generated via echo evolution (Section \ref{AppendixA2}), to a training set of $8000$ vectors, a validation set of $2000$ vectors, and a testing set of $2000$ vectors. The validation set is used for ``early stopping'' - saving parameters $(\hat{W}, \hat{b})$ of the best-performing neural network, and the testing set is used to check if the trained network can mitigate errors in unseen echo-evolution-generated data (see Section \ref{Echo_data_check}).
\begin{figure}[ht]
	\centering
        \begin{tikzpicture}[x=5.0cm,y=1.4cm]
          \message{^^JNeural network, shifted}
          \readlist\Nnod{6,7,6} 
          \readlist\Nstr{6,N_{h},6} 
          \readlist\Cstr{\strut m,h^{(\prev)},
          \bar{m}} 
          \def\yshift{0.5} 
          
          \message{^^J  Layer}
          \foreachitem \N \in \Nnod{ 
            \def\lay{\Ncnt} 
            \pgfmathsetmacro\prev{int(\Ncnt-1)} 
            \message{\lay,}
            \foreach \i [evaluate={\c=int(\i==\N && \lay==2); \y=\N/2-\i-\c*\yshift;
                         \index=(\i<\N?int(\i):"\Nstr[\lay]");
                         \x=\lay; \n=\nstyle;}] in {1,...,\N}{ 
              \node[node \n] (N\lay-\i) at (\x,\y) {$\Cstr[\lay]_{\index}$};
              
              \ifnum\lay>1 
                \foreach \j in {1,...,\Nnod[\prev]}{ 
                  \draw[connect,white,line width=1.2] (N\prev-\j) -- (N\lay-\i);
                  \draw[connect] (N\prev-\j) -- (N\lay-\i);
                }
              \fi 
              
            }
          }
          \path (N2-7) --++ (0,1+\yshift) node[midway,scale=1.5] {$\vdots$};
          
          \node[above=0.1,align=center,myorange!60!black] at (N1-1.90) {input layer\\[-0.2em]=\\[-0.2em]noisy input};
          \node[above=0.1,align=center,myblue!60!black] at (N2-1.90) {hidden layer};
          \node[above=0.1,align=center,myorange!60!black] at (N\Nnodlen-1.90) {output layer\\[-0.2em]=\\[-0.2em]denoised output};
        
        \end{tikzpicture}
	\caption{A neural network architecture which we use in this work. It has input layer dimension $6$, output layer dimension $6$, and a hidden layer dimension with $N_{h} = 200$ neurons. We use the ReLU activation function on a hidden layer output and the Tanh function on the output layer. Tanh function ensures correct spin magnetization values. $m_{j}$ denote spin vector components before neural network processing and $\bar{m}_{j}$ denote spin vector components after neural network processing.}
	\label{fig:neural_network}
\end{figure}

The trained neural network is then used to mitigate errors in forward-in-time generated data (see Section \ref{Forward_data_result} for results).
The final goal of quantum error mitigation via a neural network is to train a neural network a mapping between noisy observable data and noise-free observable data. A trained neural network thus will do a denoising map
\begin{equation}
    \vec{m}_{corrected} = f(\vec{m}_{noisy}, \hat{W}, \hat{b})
\end{equation}
for every vector $\vec{m}_{noisy}$ which we gather from the simulation of a noisy quantum device.

\subsection{Performance metric}
\label{AppendixA5}
To illustrate the efficiency of error mitigation, we introduce a correction efficiency value of the form
\begin{equation}
    \label{correction_eff_general}
    K = 1 - \frac{|\langle O\rangle_{ideal} - \langle O\rangle_{corrected}|}{|\langle O\rangle_{ideal} - \langle O\rangle_{noisy}|}
\end{equation}
Here we denote observable values for noise-free (ideal) circuit run
\begin{equation}
    \langle O\rangle_{ideal} = Tr[\hat{O}\rho_{ideal}],
\end{equation}
noisy circuit run
\begin{equation}
    \langle O\rangle_{noisy} = Tr[\hat{O}\rho_{noisy}],
\end{equation}
and noisy circuit run with error mitigation
\begin{equation}
    \langle O\rangle_{corrected} = Tr[\hat{O}\rho_{corrected}].
\end{equation}
This metric is designed to characterize effectiveness of error mitigation. If $\langle O\rangle_{corrected} = \langle O\rangle_{ideal}$, then error mitigation method completely restored the effect of quantum noise, and $ K = 1$. If $\langle O\rangle_{corrected} = \langle O\rangle_{noisy}$, then error mitigation method had no effect on the noisy observable value, and $ K = 0$. Otherwise, when $K < 0$, the error mitigation method decreased the quality of observable value. 

In this work, we used average system magnetization as an observable of interest. For average magnetization, the correction efficiency is
\begin{equation}
    \label{correction_eff}
    K =  1 - \frac{|\Delta M_{after}|}{|\Delta M_{before}|}
\end{equation}
where
\begin{eqnarray}
    \label{DeltaM}
    \Delta M_{before} = M_{ideal} - M_{noisy}\\
    \Delta M_{after} = M_{ideal} - M_{corrected}, 
\end{eqnarray}
where $M_{ideal}$, $M_{noisy}$ and $M_{corrected}$ are defined as (\ref{average_m}).
This coefficient allows illustrating the effect of error mitigation via neural network, comparing resulting observables: if $\Delta M_{after} < \Delta M_{before}$, then the observable error reduced after neural network processing; in opposite case, when $\Delta M_{after} > \Delta M_{before}$, the observable error increased after neural network processing.

\section{Results}
\label{section4}

\subsection{Quantum error mitigation in forward-in-time evolution}
\label{Forward_data_result}

\begin{figure}[ht]
	\centering
	\includegraphics[width=0.99\linewidth]{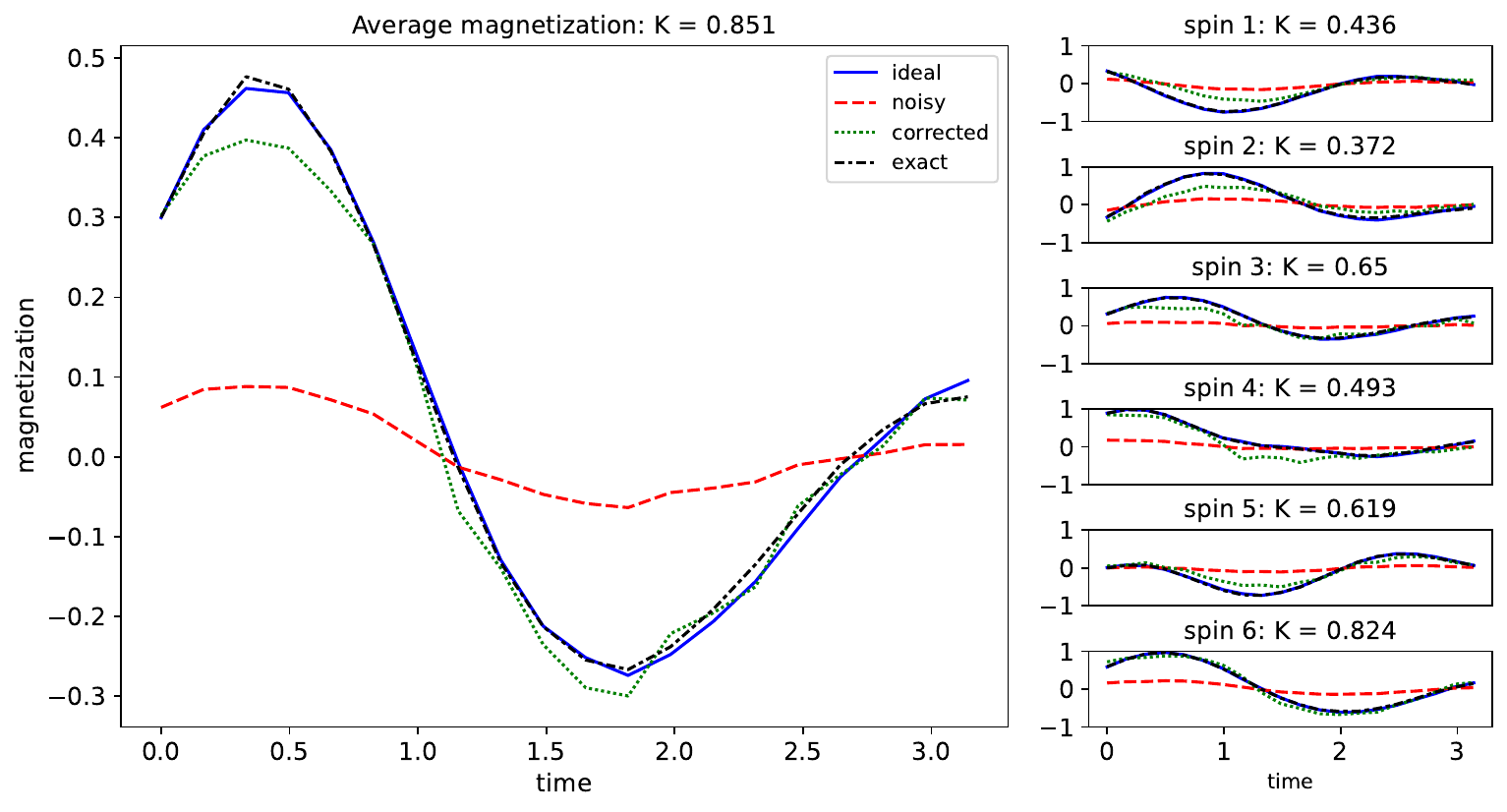}
	\caption{
 \textbf{Left:} forward-in-time evolution of average chain magnetization of a random initial state with noise-free evolution (blue), noisy evolution (red), corrected with neural network (green), and exact evolution (black). We simulate noise-free evolution, noisy evolution, and neural-network-corrected evolution via Trotterization of a hamiltonian evolution with the same number of Trotter steps ($N_{tr} = 20$), and simulate ``exact'' dynamics with $10 N_{tr}$ Trotter steps.
 \textbf{Right:} Forward-in-time evolution of magnetization of individual spins for the initial state, described on the left plot.}
	\label{fig:main_forward_fig1}
\end{figure}

\begin{figure}[ht]
	\centering
	\includegraphics[width=0.9\linewidth]{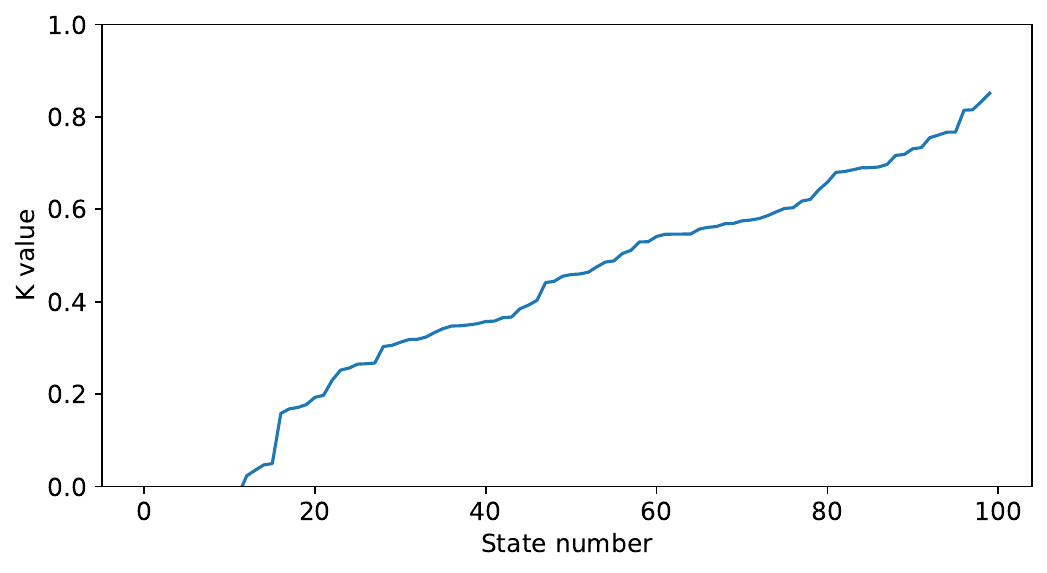}
        \caption{Values of correction efficiency $K$ for forward-in-time evolution of 100 random initial states. States are sorted with respect to the correction efficiency value $K$. Every $K$ value is calculated for average magnetization values of magnetization of the spin system, with further averaging over $20$ time points to produce a single value (i.e., average over 20 $K$ values for each time point).}
	\label{fig:main_forward_fig2}
\end{figure}

Here, we use a neural network, trained on an echo dynamical data set, to correct forward-in-time evolution results under the transverse field Ising hamiltonian (\ref{Ising_hamiltonian}). The test data set consists of $100$ vectors for different initial states (see Appendix \ref{AppendixA} for simulation details). For every initial state, we simulated noisy dynamics for $20$ time points in the range $[0, \pi]$ (in $h$ units) and calculated the evolution of single spin magnetizations and average magnetization over the spin system. We provide an example of the corrected evolution of magnetization with a trained neural network (for both single spins and average magnetization) in Fig.~\ref{fig:main_forward_fig1}.
To characterize correction efficiency over different initial states of a spin system, we introduce a correction efficiency value $K$. If $K = 1$, the correction corresponds to a perfect correction, $K = 0$ corresponds to no correction, and $K < 0$ corresponds to a decrease of results quality after neural network post-processing (see Section \ref{AppendixA5} for details).
In Fig.~\ref{fig:main_forward_fig2}, we provide correction efficiency value $K$ for forward dynamics of 100 initial random states. The provided $K$ values for every state are calculated for average values of magnetization of the spin system and are further averaged over all 20 time points to produce a single value (i.e., average over 20 $K$ values for each time point).

We can see that a neural network trained on echo evolution can correct forward-in-time evolution results. 
Thus, a neural network trained to mitigate the quantum noise effect on echo evolution can mitigate the quantum noise effect in forward-in-time evolution. 
The quality of correction for average magnetization dynamics is predictably better than for single spin magnetization (see $K$ values for average (left) and single spin (right) dynamics in Fig.~\ref{fig:main_forward_fig1}). 
From Fig.~\ref{fig:main_forward_fig2}, we see that most of the states from a 100-state sample have a positive correction efficiency value $K$.

\begin{figure}[ht]
	\centering
	\includegraphics[width=1.2\linewidth]{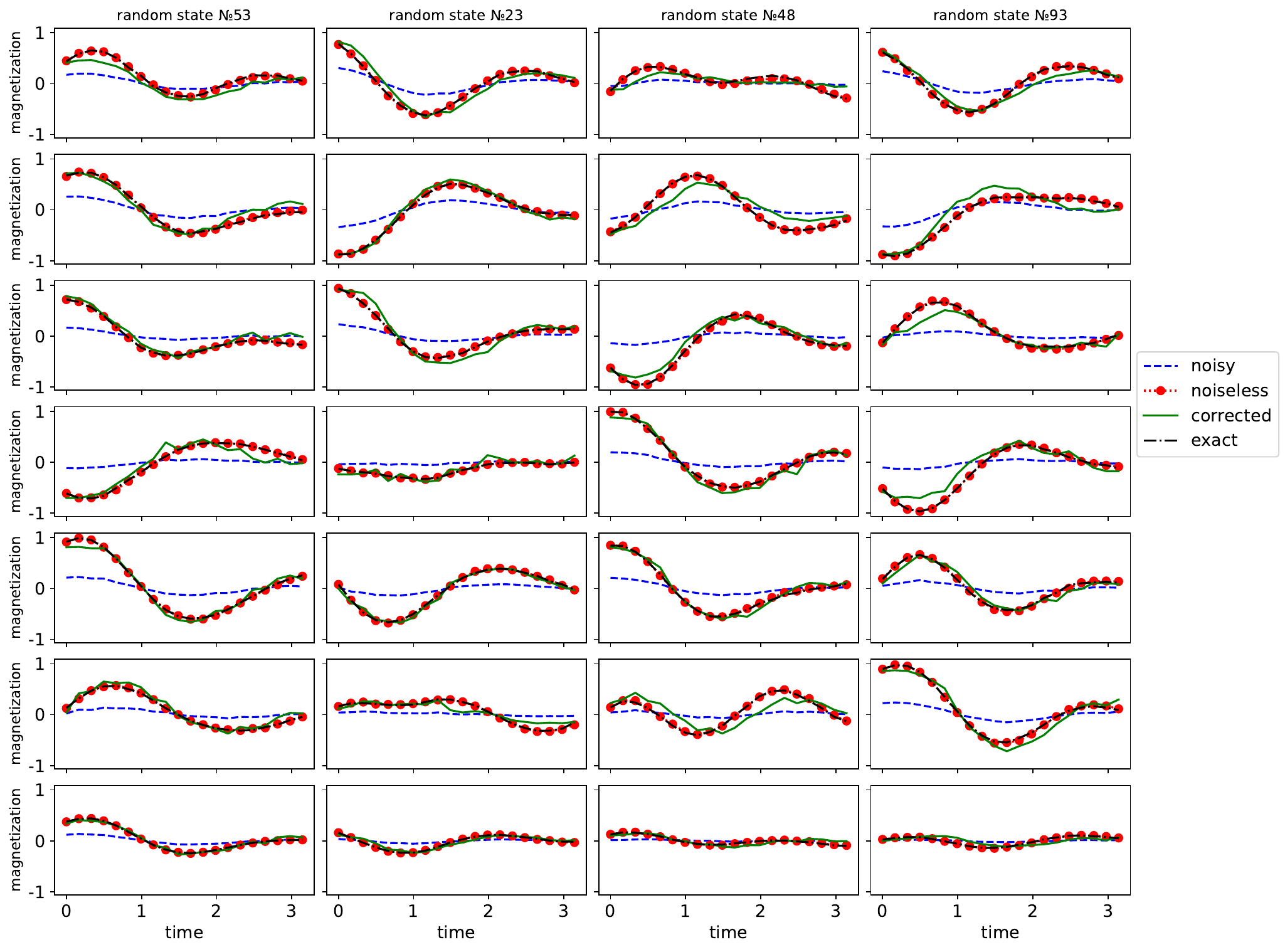}
	\caption{Forward-in-time evolution of individual spin magnetizations for 4 random initial states.}
	\label{fig:main_forward_fig3}
\end{figure}

Finally, in Fig.~\ref{fig:main_forward_fig3}, we provide examples of single spin magnetization dynamics before and after neural network correction. We again see that a neural network can mitigate the effect of quantum gate noise on the quality of observable dynamics of the forward-in-time evolution.

An interesting point is that the proportion of data with positive error mitigation ($K > 0$) is approximately $88\%$. 
The same proportion of corrected states is observed, if we correct echo evolution instead of forward evolution. In particular, if we generate data vectors (vectors of spins magnetization) and divide the whole data set into a training part and a test part, applying a trained neural network to a test part - which was not used in the training process - will give a positive correction efficiency proportion similar to what we observe in the case of correcting forward-in-time evolution. More analysis on the correction efficiency proportion in case of echo evolution test data is provided in Appendix \ref{Echo_data_check}. 
This proportion can be improved with doing more shots for every quantum circuit.
It is unclear how this proportion depends on the noise level of quantum gates. Experiments with another level of noise (done by authors during this work) do not allow corroboration of any hypothesis.


\subsection{Analysis of the neural network size}
\label{AnalysisWidth}

In this section, we demonstrate that, for the dynamics we here consider, the number of neurons in the hidden layer can be equal to approximately the size of data vectors. 
For that purpose, we train ensembles of neural networks with different widths of their hidden layers on different subsets of echo-evolution-generated data. Then, we calculate correction efficiency (\ref{correction_eff}) and statistics of changes in magnetization data vectors before and after postprocessing with a trained neural network (\ref{DeltaM}). We keep all other hyperparameters fixed (see Appendix \ref{AppendixNNsize} for details).

In Fig.~\ref{fig:echo_K_vals}, we provide dependence of the correction efficiency value (\ref{correction_eff}) on the width of the hidden layer for different levels of quantum noise (values of two-qubit gates error $q_{2}$).
Each value is calculated for average magnetization values, and the statistics in calculated over 50 different neural networks for each width of the hidden layer and for 1000 initial states in each training set (different for every of 15 neural network hidden layer width values.)
From Fig.~\ref{fig:echo_K_vals}, we see that the correction efficiency (\ref{correction_eff}) for noisy data stops increasing after approximately 8-neurons width of the hidden layer - even for 200 neurons in the hidden layer, we have almost the same performance of error mitigation as we have with an 8-neuron layer. 
We provide additional details on statistics of average magnetization differences in Appendix \ref{AppendixNNsize}.

This result is surprising since making neural networks larger generally leads to better results. We see that a relatively small width of the hidden layer was enough 
to saturate the neural network quantum error mitigation performance.  
Here, we investigated a single setup - the dynamics of a spin system with a fixed size, implemented with only a particular kind of single- and two-qubit gate noise (depolarizing). It is interesting to investigate other types of quantum noise and more systems with different quantum observables.

\begin{figure}[ht]
	\centering
	\includegraphics[width=0.8\linewidth]{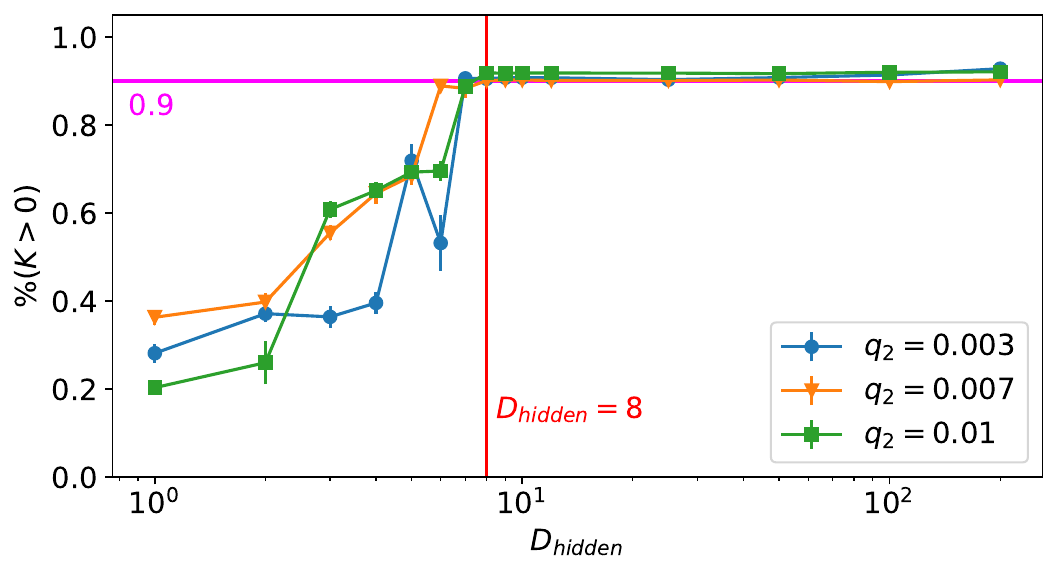}
	\caption{Values of correction efficiency (\ref{correction_eff}) for neural networks with different widths of the hidden layer. Each value is calculated for average magnetization values, and the statistics in calculated over 50 different neural networks for each width of the hidden layer and for 1000 initial states in each training set (different for every of 15 neural network hidden layer width values.) Results provided for data with different levels of noise (indicated with different values of two-qubit gate noise $q_{2}$).}
	\label{fig:echo_K_vals}
\end{figure}

\section{Conclusion}
\label{Conclusion}

We introduced a physics-motivated method to generate data for training neural networks for quantum error mitigation. This method uses the echo evolution of a quantum system to generate noisy and noise-free data vectors.  
This method does not require classical simulation and simplification of target evolution circuit and thus is practically applicable to problems of the quantum advantage size($\geq 50$ qubits).
We demonstrated that a neural network trained on echo-evolution-generated data mitigates effect of quantum noise on results of the forward-in-time evolution.
For illustration, we used a system of 6 spins, evolving under the transverse-field Ising Hamiltonian, whose evolution we implemented via noisy quantum gates. 
As a side result, we observed that the width of one hidden layer of our network only requires about ten neurons to 
saturate neural network performance in quantum error mitigation.

There are several ways to improve the proposed method. The method uses random initial states to generate data set with echo evolution. In this form, the method intrinsically uses the fact that the depth of circuits used to prepare initial states is much more shallow than evolution-generating circuits. Although it is practically reasonable, it goes with a certain amount of noise, which affects the noise-free part of the data. The first possible solution is reducing the noise by limiting the initial states to states generated with depth-restricted circuits. For example, we can use only one layer of single-qubit rotations to generate initial states. As provided in \cite{Kim2022, Lee2023}, one layer of single-qubit rotations allows the generation of almost noise-free states with exactly known form and, thus, with known observable values. 
The second possible solution is to use Clifford gates to prepare initial states, restricting the state-preparing circuit depth to shallow (i.e., much less depth that the evolution circuit). In that case we still have access to the ideal observable value and prepare initial states with negligible error with respect to the forward evolution (which is non-Clifford).

The method we provide here can stimulate applications of neural networks for quantum error mitigation. 
Data-driven QEM provides a prospective tool to enhance the capabilities of NISQ devices. 
In general, all data-driven methods learn a mapping between noisy and noise-free observable data. 
Recent results provide reasoning that, for large quantum circuits, this mapping could be of almost rescaling form \cite{Babukhin2021,tsubouchi2023universal}, as the cumulative quantum noise acts almost as white noise \cite{dalzell2021random}.
So, in essence, data-driven methods provide a clever way of learning how to rescale (with a slight nonlinearity) observable values affected by quantum noise.

We have already seen that neural networks can mitigate quantum evolution noise \cite{Zhukov2022} and
quantum measurement noise \cite{Kim2022,Lee2023}.
At the same time, mitigating coherent errors (e.g., gate angle over-rotation, phase calibration errors, crosstalk-induced evolution) with machine learning is still a scarcely explored field. Recent works demonstrated that coherent errors severely reduce mitigation efficiency of machine learning methods \cite{Zhukov_2024, liao2023machinelearningpracticalquantum}.
A possible solution for that problem was proposed to reduce the effect of phase calibration errors and gate angle over-rotation \cite{Zhang_2022,kaufmann2023characterizationcoherenterrorsnoisy,Majumder_2023}. In this work, coherent errors are mitigated via alternately applying direct and ``inverse'' sequences of control pulses, which implement the same quantum gate (conditions when it is possible are also provided). While this method reduces phase calibration and over-rotation errors, a crosstalk-induced evolution cannot be mitigated this way as crosstalk error depends on qubit excitation and circuit depth rather than on parameters of gate pulses. Possible ways to do quantum error mitigation in the presence of crosstalk error is to use randomized compiling \cite{Wallman_2016,Perrin_2024} or apply dynamical decoupling sequences \cite{niu2024,evert2024}. As the effect of crosstalk-induced dynamics depends circuit depth, it would be interesting to incorporate circuit depth as an additional dimension in data vectors. Whether it allows to mitigate crosstalk-induced errors is a subject of future research.

As neural networks enter the field of quantum error mitigation, an important question is to compare neural networks to other methods \cite{Cai2023}. In particular, the scaling of neural network size and efficiency concerning various experiment variables (e.g., data dimension, training sample size, noise level, or neural network width/depth) is unknown. Other QEM techniques (e.g., Zero Noise Extrapolation and Probabilistic Error Cancellation \cite{Temme2017}, Virtual Distillation \cite{Huggins2021}) have known efficiency under a general framework \cite{Cai2023practical}.
As we demonstrated, a neural network with a single hidden layer can saturate its quality with a small hidden layer width. It is possible that the number of neurons needed to mitigate quantum noise effect depends on the complexity of the system dynamics. The spin dynamics considered in this work required a number of neurons, which is equal to approximately the size of the system.
Further experiments with systems of different sizes and types of quantum noise and theoretical scaling analysis are interesting subjects for future research.

\backmatter

\bmhead{Acknowledgements}
The author would like to thank W.V. Pogosov for careful reading the manuscript and providing feedback. Simulation were made using qiskit library \cite{Qiskit}. Illustrations for quantum circuits were made using quantikz library \cite{kay2023tutorial}. 

\bmhead{Data availability statement} 
All code and data, supporting this manuscript, are available in github repository \cite{git_ref}.

\bmhead{Note added} 
After the completion of this manuscript, a work about generation data for neural network quantum error mitigation appeared \cite{liao2023}, where authors propose a framework of fiducial processes, which also explores generation noisy and noise-free data without classical simulation of the target quantum computation process. The echo-evolution data generation can be considered an example of this framework application, inspired by process known in dynamics of physical systems. Although we here concentrated on dynamics of spin systems, echo-evolution can be applied to all processes, based on hamiltonian simulation.

\section*{Declarations}

\bmhead{Conflicts of interest}

Author declares no conflict of interest.









\begin{appendices}

\section{Supervised learning of feed-forward fully-connected neural networks}
\label{supervisedlearning}

In essence, neural networks are composite nonlinear functions which can approximate a data-underlying function. Recently, neural networks were used as an effective instrument for solving problems in physics \cite{Mehta2019}. Eventually, the use of neural networks spread to quantum informatics \cite{Palmieri2020,Neugebauer2020,Bukov2018} and to quantum error mitigation \cite{Kim2020,Kim2022,Zhukov2022,Kim2023}. It is a reasonable advancement of applications since the error mitigation aims to post-process data from a noisy quantum device with an approximately inverse noise map, which can be a function that the neural network approximates. Here, we provide some notions of neural network machinery we will need in the following.

The most straightforward approach to training neural networks is supervised learning. To formulate this approach, we need to recall a notion of the data set. Suppose we have vectors $\vec{x}$ and corresponding ``labels'' $\textbf{y}$. 
A set of these pairs $\{\vec{x}_{i}, \textbf{y}_{i}\}|_{i=1}^{N_{data}}$ belongs to a population of all vectors $\vec{x}$ and $\textbf{y}$ and have an underlying functional dependency $\mathcal{F}(\vec{x}) = \textbf{y}$. The set $\{\vec{x}_{i}, \textbf{y}_{i}\}|_{i=1}^{N_{data}}$ is called a data set. Depending on structure of $\textbf{y}$, the data set can induce a classification problem ($\textbf{y} \in [1, 2, ...]$), regression problem ($\textbf{y} \in \mathbb{R}$), or be another kind of mapping problem ($\textbf{y} = \vec{y}$).

The aim of supervised learning is to train a parameterized function $f(\vec{x}, \hat{W})$ - for example, a neural network - to approximate the function $\mathcal{F}$ on the whole domain of values, using only available data set $\{\vec{x}_{i}, \textbf{y}_{i}\}|_{i=1}^{N_{data}}$. Such a neural network must have many parameters and enough non-linearity to be expressive to learn a data function $\mathcal{F}$ from provided data. The most simple architecture of a neural network is a fully connected neural network. Every layer of such a network does a nonlinear map of the form
\begin{equation}
    x_{out}^{j} = \sigma
    \biggl(
    \sum_{i=1}^{N_{in}}W_{ji}x_{in}^{i} + b_{j}
    \biggl), 
    \hspace{0.25cm} 
    \vec{x}_{out} = (x_{out}^{1}, x_{out}^{2}, ..., x_{out}^{N_{out}}).
\end{equation}
Here, $W_{ji}$ and $b_{j}$ are neural network parameters to be trained, and $\sigma(.)$ is a non-linear function, $x_{in}^{i}$ and $x_{out}^{j}$ are vector components of input and output vectors, $N_{in}$ and $N_{out}$ - number of elements in $x_{in}$ and $x_{out}$.
Several such layers compose a neural network. For example, a neural network with an input layer, a single hidden layer, and an output layer has the following form:
\begin{equation}
    \label{neuralnet_func}
    f(\vec{x}, \hat{W}, \hat{b}) 
    = 
    \sigma_{2}
    \biggl(
    \sum_{j_{2}=1}^{N_{2}}W_{j_{2}j_{1}}
    \sigma_{1}
    \biggl(
    \sum_{j_{2}=1}^{N_{1}}W_{j_{1}i}x_{in}^{i} + b_{j_{1}}
    \biggl)
    + b_{j_{2}}
    \biggl).
\end{equation}
Here we denoted all trainable weights on different layers as $\hat{W}$ and all trainable biases as $\hat{b}$, $\sigma_{1(2)}$ - non-linear functions, applied after hidden (output) layer correspondingly.
For our purposes, the understanding of neural networks as functions of the form (\ref{neuralnet_func}), which takes vectors $\vec{x}_{in}$ as inputs and returns vectors $\vec{x}_{out} = f(\vec{x}_{in}, \hat{W}, \hat{b})$ as outputs, is sufficient.

Having data set $\{\vec{x}_{i}, \textbf{y}_{i}\}|_{i=1}^{N_{data}}$ and a neural network $f(\vec{x}, \hat{W}, \hat{b})$, the goal is to train a neural network to approximate functional dependence of data set such that the network can realize this dependence on unseen data vectors $\vec{x}$. The training usually requires two ingredients - choosing a loss function and a training optimization procedure. The loss function $L(f(\vec{x}, \hat{W}, \hat{b}), y)$ measures the error between the network output and a correct value of $\textbf{y}$. Calculating loss values over the data set, we construct an empirical risk
\begin{eqnarray}
    \label{empirical_risk}
    R(\hat{W}, \hat{b}) = \frac{1}{N_{data}} \sum_{i=1}^{N}L(f(\vec{x}_{i}, \hat{W}, \hat{b}), \vec{y}_{i})
\end{eqnarray}
which we minimize over iterations of training. The training process is a gradient descent towards the minimum of empirical risk (\ref{empirical_risk}), with the use of back-propagation of error through weights in different layers \cite{nielsenneural}. Usually, the training process continues until a satisfactory local minimum of the function (\ref{empirical_risk}) is reached and a set of neural network parameters is known
\begin{equation}
    (\hat{W}_{*}, \hat{b}_{*}) = arg \min_{W, b} R(\hat{W}, \hat{b})
\end{equation}

\section{Quantum error mitigation in echo evolution data}
\label{Echo_data_check}
We train a neural network on echo-evolution-generated data and check its capability to correct data of the same kind, i.e., generated with echo evolution. We use the experiment setup described in Appendix \ref{AppendixA}.
In Fig.~\ref{fig:main_echo_fig1}, we provide statistics of average magnetization error for 2000 test vectors. 
In particular, in the test data set, we have 2000 pairs of noisy and noise-free vectors of spin magnetizations of the form $(\vec{m}_{noisy}, \vec{m}_{ideal})$ (see \ref{magnetization_vect}). From noise vectors we obtain, after applying a trained neural network for data vectors $\vec{m}_{noisy}$, 2000 corrected vectors $\vec{m}_{corrected}$. Then we calculate the differences between the average magnetization of the spin system, calculated from noise-free and noisy vectors, and noise-free and corrected vectors:
\begin{equation}
    \label{deltaM}
    \Delta M_{noisy(corrected)} = \frac{1}{N}\sum_{j}(m^{j}_{ideal} - m^{j}_{noisy(corrected)}),
\end{equation}
where the second vector is a noisy (neural network corrected) vector of spin magnetization. Values of $\Delta M$ are provided in Fig.~\ref{fig:main_echo_fig1} (left plot), and absolute values of $\Delta M$ for every one of 2000 test states are provided in Fig.~\ref{fig:main_echo_fig1} (right plot).
We see that applying a neural network leads to decreased error dispersion (left histogram) and decreased absolute error value (right scatter plot).
\begin{figure}[ht]
	\centering
	\includegraphics[width=1.0\linewidth] 
             {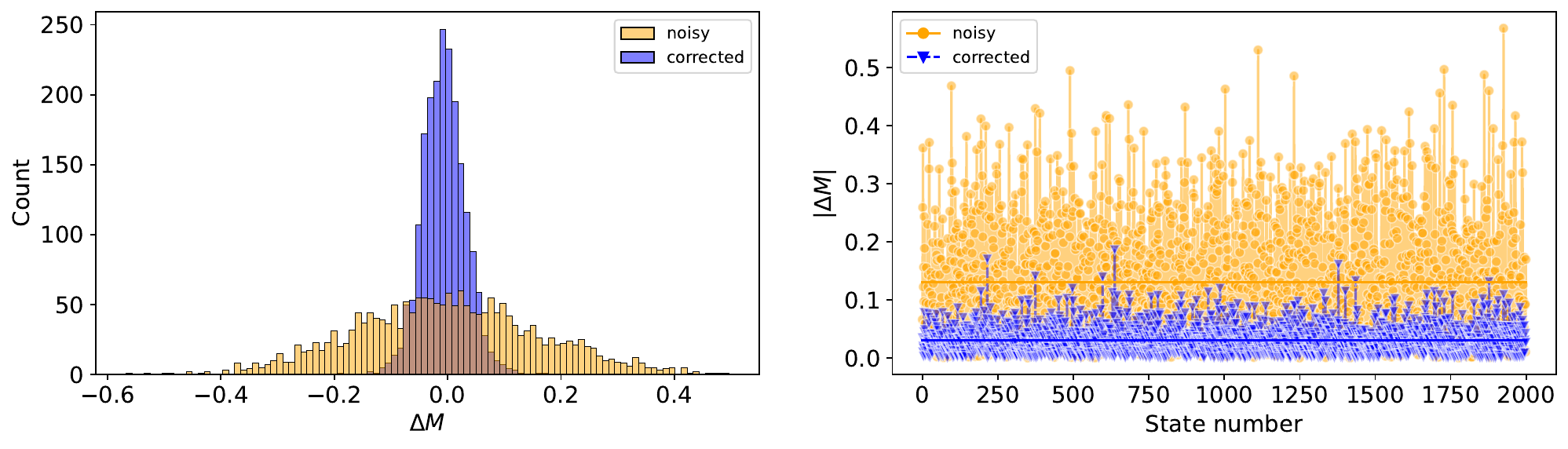}
	\caption{
 \textbf{Left:} a histogram of $\Delta M$ distribution of echo evolution results before and after error mitigation with a trained neural network. 
 \textbf{Right:} a scatter plot of $\Delta M$ for 2000 final states after echo evolution before and after error mitigation with a trained neural network. Solid lines denote average absolute values of $\Delta M$ over the test vectors dataset.}
	\label{fig:main_echo_fig1}
\end{figure}
We provide correction efficiency values $K$ (see (\ref{correction_eff})) for every test data vector alongside with ideal, noisy, and corrected average magnetization of the spin system in Fig.~\ref{fig:main_echo_fig2} (we provide noise-free/noisy and corrected magnetization values on two separate figures for better visibility without overlapping). Here, magnetization values are sorted with respect to correction efficiency value $K$, thus there is an ascending character of $K$ value with state number in Fig.~\ref{fig:main_echo_fig2}.  
We can see that the neural network compensates for the shrinkage of average magnetization for most of the states in the test sample: approximately $88\%$ of states have a positive correction efficiency value $K$, which means that, after applying the neural network, the difference between corrected and ideal magnetization values decreased with respect to the difference between uncorrected and noise-free values. We note that states that are hard to correct with a neural network (negative values of $K$) have average magnetization close to zero. As depolarizing gate error leads to a decrease in the average magnetization absolute value (it tends to zero), if the ideal state has close to zero average system magnetization, its noise-corrupted magnetization is indistinguishable from the noise-free case.  
\begin{figure}[ht]
	\centering
	\includegraphics[width=1.0\linewidth]{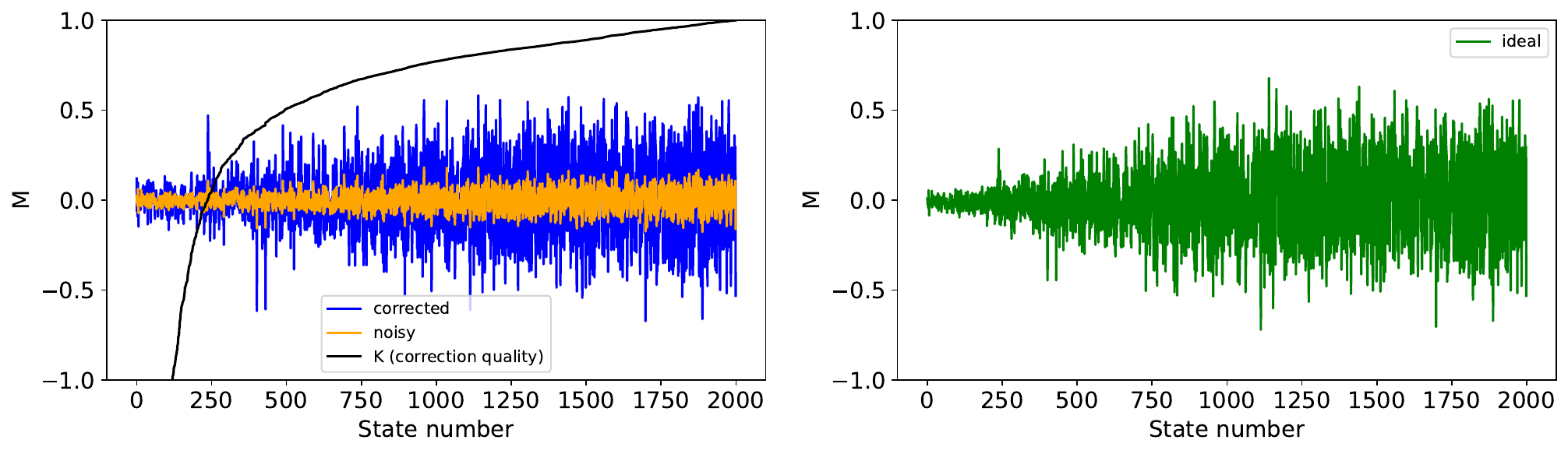}
	\caption{
 \textbf{Left:} magnetization values (averaged over the spin chain) for 2000 final states after echo evolution with noisy evolution operator (orange) and after error mitigation with a trained neural network (blue). Values are sorted with respect to the correction efficiency parameter $K$ (black curve).
 \textbf{Right:} magnetization values (averaged over the spin chain) for 2000 final states after echo evolution with 
 exact evolution operator. Values are sorted with respect to the correction efficiency parameter $K$.}
	\label{fig:main_echo_fig2}
\end{figure}

\section{Performance under different noise levels}
\label{AppendixNoise}
Here we provide additional data on neural network quantum error mitigation under different levels of gate error noise. We executed quantum error mitigation with a neural networks, trained on echo-evolution generated data for several levels of gate noise for 100 forward-in-time evolution of random initial states. We calculated correction efficiency values $K$ for every evolution as well as percentage of positive $K$ values in each initial states sample.
\begin{figure}[ht]
	\centering
	\includegraphics[width=0.9\linewidth]{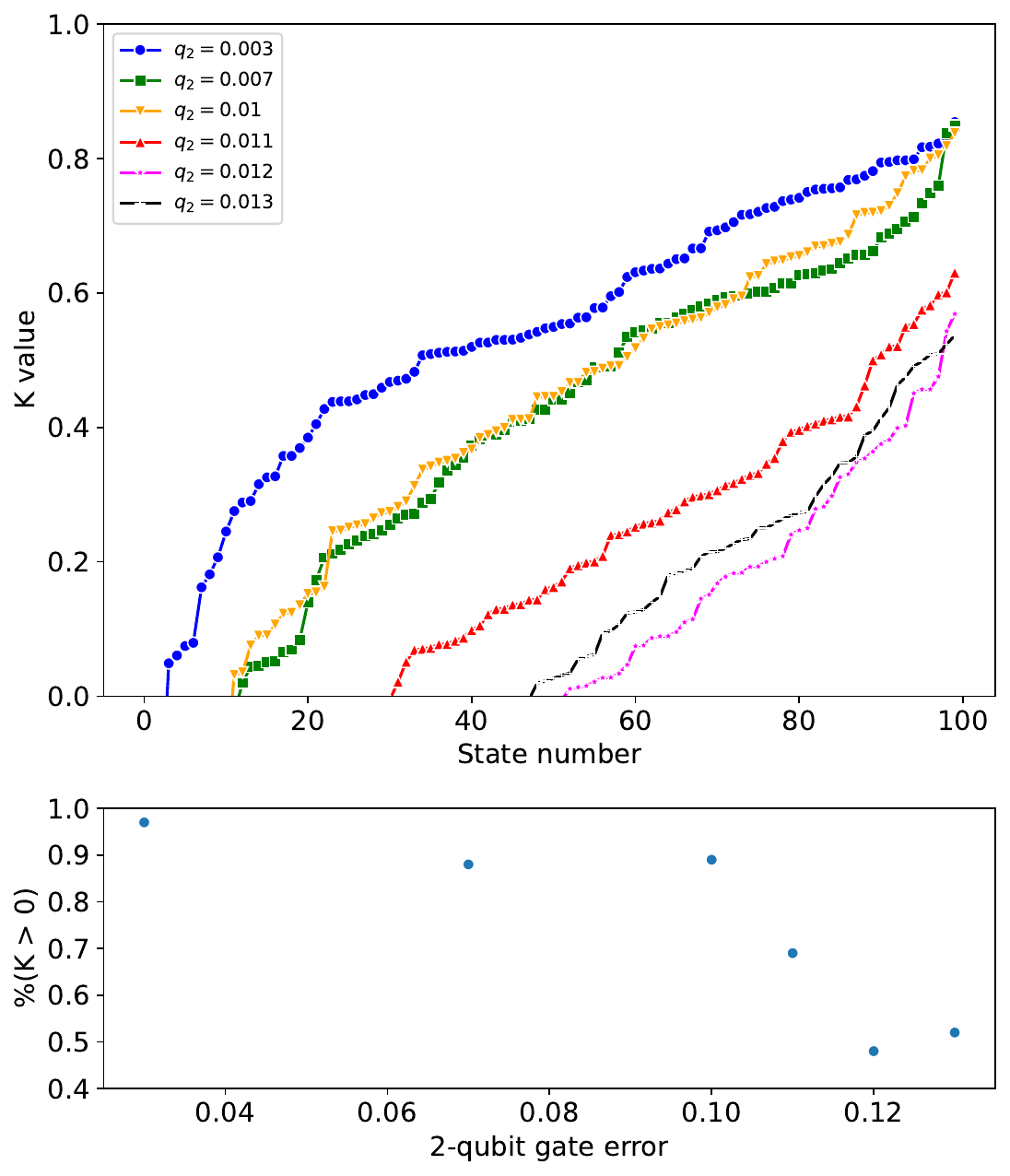}
        \caption{
        \textbf{Upper}: Values of correction efficiency $K$ for forward-in-time evolution of 100 random initial states for different gate errors (two-qubit gate errors $q_{2}$ are provided to quantify noise power). States are sorted with respect to the correction efficiency value $K$. Every $K$ value is calculated for average magnetization values of magnetization of the spin system, with further averaging over $20$ time points to produce a single value (i.e., average over 20 $K$ values for each time point).
        \textbf{Lower}: percentage of positive $K$ values for every initial states sample with 100 states.
        }
	\label{fig:noises_forward}
\end{figure}

We provide plots for calculation results in Fig.~\ref{fig:noises_forward}. We see that performance of correction decreases with growing power of gate noise. For noise level (per gate) below $1\%$ we see that correction efficiency values decreased down to $50\%$ of positive correction efficiency values in a sample. Although we provide here only a single sample of initial states forward-in-time evolution for each noise level, we can conclude that with more noise training a neural network to quantum error mitigation will require more efforts. Increasing size of the training data set can possibly be a requirement
to increase error mitigation efficiency of a neural network.

\section{Analysis of the neural network size, continued}
\label{AppendixNNsize}

In the main text, we trained a neural network with a single hidden layer on data obtained from the echo evolution of the spin system. Then we demonstrated quantum error mitigation in data obtained from the forward-in-time dynamics of the spin system. For that experiment, we chose a particular size of the hidden layer (200 neurons). 
In this section, 
we provide analysis illustrating that the number of neurons in the hidden layer, in the case of dynamics simulation considered here, can be lowered to approximately the size of data vectors.
In the following, we provide simulations for neural networks with variable width of the hidden layer and with other parts of the setup fixed - e.g., for the number of hidden layers, activation functions, loss function, and training procedure details (type of optimizer, size of data batches, etc.). 

To analyze this question, we evaluate the quality of error mitigation depending on hidden layer width.
In particular, we use a trained neural network to mitigate quantum noise error in the test data set - a hold-out part of echo evolution data to evaluate error mitigation performance. For a test set of a fixed size, we can calculate statistics on error-mitigated observable values and thus see how good our trained network performs error mitigation. 

We generated via echo evolution data sets (12000 pairs of noise and noise-free magnetization vectors) for different levels of two-qubit gate noise ($q_{2} = 0.003, 0.007, 0.01$). For every noise level, we did the following steps. 
First, we generated $50$ random subsets of training data (6000 pairs of noise and noise-free magnetization vectors) and divided each subset into a training data set (4000 vector pairs), validation data set (1000 vector pairs), and test data set (1000 vector pairs). 
Second, we chose a width of the hidden layer $N_{hidden} \in [1, 2, 3, 4, 5, 6, 7, 8, 9, 10, 12, 25, 50, 100, 200]$ and generated 50 neural network models with the same initialization of weights. 
Third, we trained these neural network ``clones'' on the 50 different data sets, each for 100 epochs. During the training, we used a batch size of 80, the Adam optimizer with learning rate $lr = 3*10^{-4}$, $\beta_{1} = 0.9$, $\beta_{2} = 0.999)$.
We used mean square error (\ref{MSE}) as a loss function.
Finally, we calculated average correction efficiency values and several statistics of $|\Delta M|$ - difference of average magnetization of spin system before and after applying a trained neural network (see \ref{DeltaM}).
The average was calculated over the 50 different realizations of neural networks with fixed hidden layer width. 
Similarly, correction efficiency values were calculated for the 50 different test sets with 1000 vector pairs, and an average of over these 1000 vectors generated a single value for each test data set.

\begin{figure}[ht]
	\centering
	\includegraphics[width=0.99\linewidth]                       {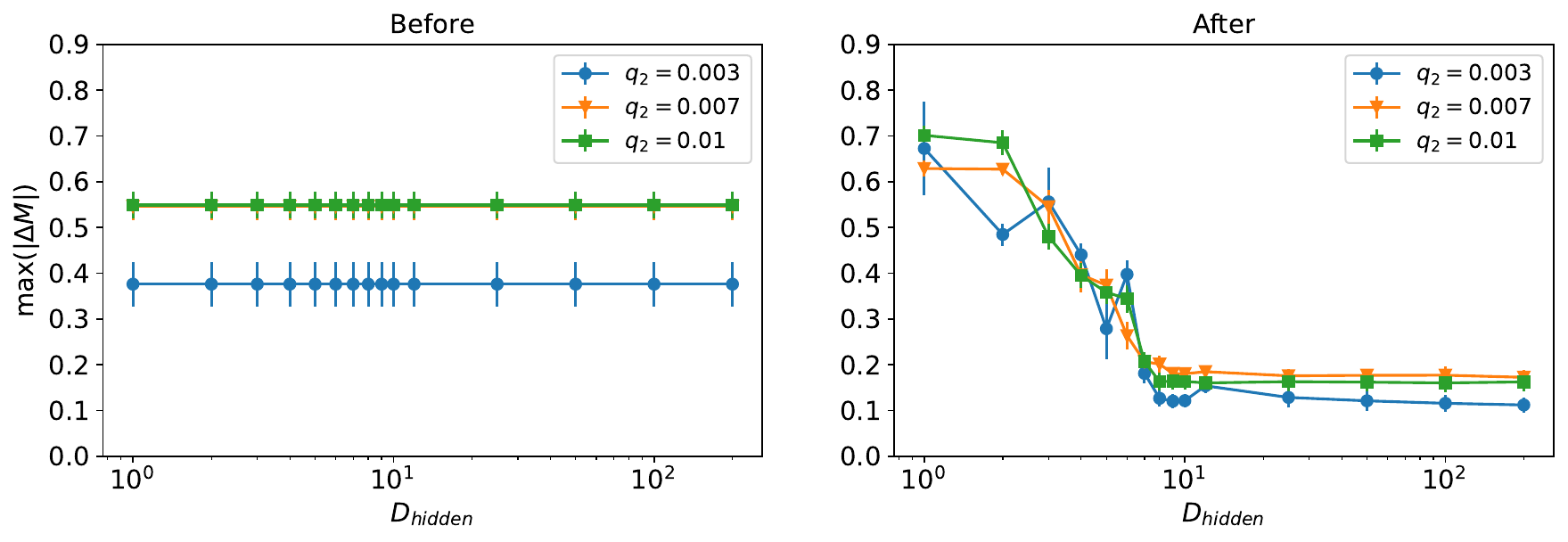}
        \includegraphics[width=0.99\linewidth]{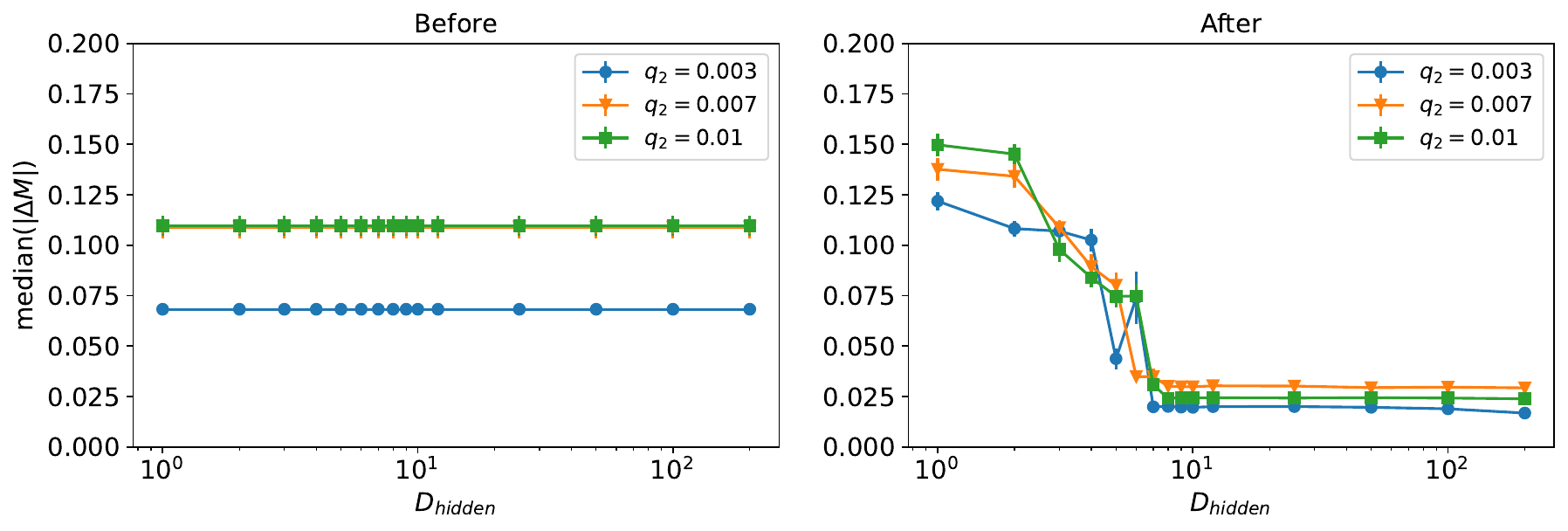}
        \includegraphics[width=0.99\linewidth]{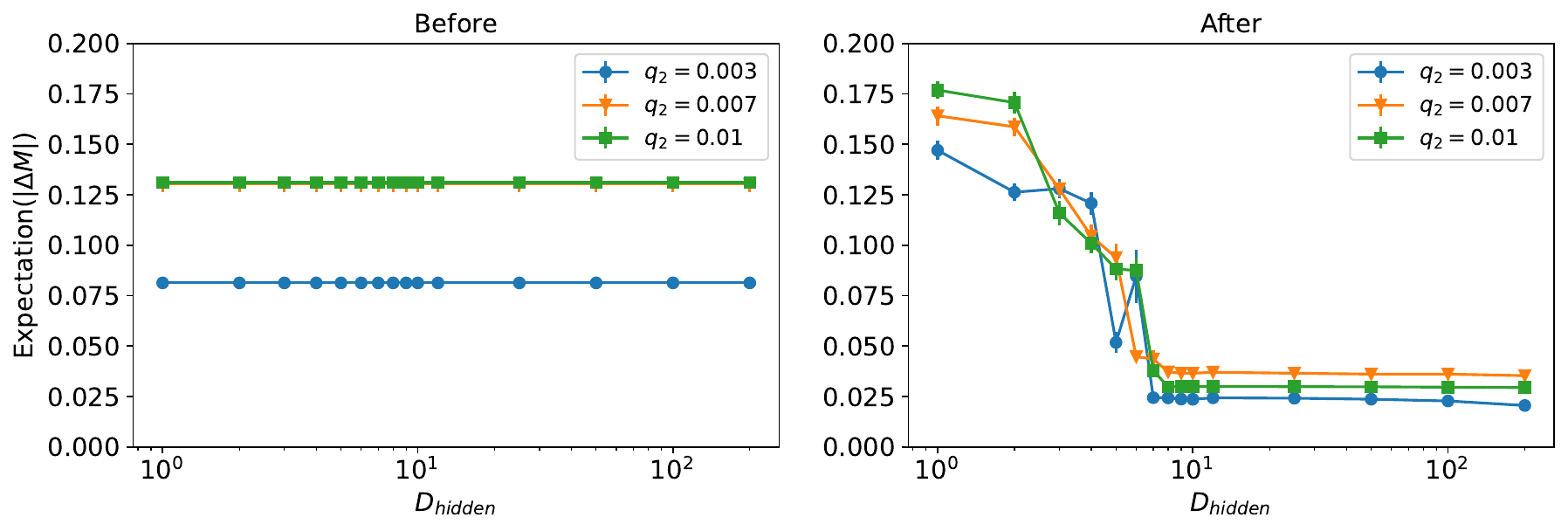}
	\caption{Values of observable values statistics for neural networks with different widths of the hidden layer. Average values and standard deviation are calculated over results from 50 realizations of neural networks with fixed hidden layer width. Results are provided for data with different levels of noise (indicated with different values of two-qubit gate noise $q_{2}$). \textbf{Left:} observable values statistics before correction with a trained neural network. \textbf{Right:} observable values statistics after correction with a trained neural network.}
	\label{fig:echo_statistic_vals}
\end{figure}

In the main text in Fig.~\ref{fig:echo_K_vals}, we provided dependence of the correction efficiency (\ref{correction_eff}) on the width of the hidden layer for different levels of quantum noise (indicated with different values of two-qubit gates error $q_{2}$). 
Here, in Fig.~\ref{fig:echo_statistic_vals}, we provide dependence of average magnetization difference (\ref{deltaM}) on the hidden layer width.
From Fig.~\ref{fig:echo_statistic_vals} we can see (similar to Fig.~\ref{fig:echo_K_vals}) that the performance of quantum error mitigation becomes saturated with the hidden layer width of approximately 8 neurons. This pattern also occurs for different levels of quantum noise. 

\end{appendices}


\bibliography{ref}

\end{document}